\documentclass[%
 aip,
 amsmath,amssymb,floatfix,
 reprint%
]{revtex4-1}\usepackage[utf8]{inputenc}
\usepackage{CJK}
\usepackage[english]{babel}

\usepackage{graphicx}
\usepackage{xcolor}
\usepackage{natbib}
\usepackage{amsmath}
\usepackage{amssymb}
\usepackage{bm}
\usepackage{ar}

\newcommand{\etal}{et al.\ }
\newcommand{\ie}{i.e.,\ }

\newcommand{\secref}[1]{\mbox{section \ref{#1}}}

\newcommand{\equref}[1]{\mbox{equation (\ref{#1})}}

\newcommand{\figrefC}[2][]{\mbox{Figure \ref{#2}(#1)}}
\newcommand{\figref}[2][]{\mbox{figure \ref{#2}(#1)}}
\newcommand{\figrefSC}[1]{\mbox{Figure \ref{#1}}}
\newcommand{\figrefS}[1]{\mbox{figure \ref{#1}}}

\newcommand{\mr}[1]{{\color{black}#1}}

\begin{document}
\title{Shear-thinning and shear-thickening emulsions in shear flows}
%高木 周
\author{Marco E. Rosti}
\email[Corresponding author: ]{marco.rosti@oist.jp}
\affiliation{Complex Fluids and Flows Unit, Okinawa Institute of Science and Technology Graduate University, 1919-1 Tancha, Onna-son, Okinawa 904-0495, Japan}
\affiliation{Department of Mechanical Engineering, The University of Tokyo, Tokyo, Japan}

\author{Shu Takagi \begin{CJK*}{UTF8}{gbsn}(高木 周)\end{CJK*}}
\affiliation{Department of Mechanical Engineering, The University of Tokyo, Tokyo, Japan}

\begin{abstract}
We study the rheology of a two-fluid emulsion in semi-concentrated conditions; the solute is Newtonian while the solvent an inelastic power law fluid. The problem at hand is tackled by means of direct numerical simulations using the volume of fluid method. The analysis is performed for different volume fractions and viscosity ratios under the assumption of negligible inertia and zero buoyancy force. Several carrier fluids are considered encompassing both the shear-thinning and thickening behaviors. We show that the effective viscosity of the system increases for shear-thickening fluids and decreases for the shear-thinning ones for all the viscosity ratios considered. The changes in the emulsion viscosity are mainly due to modifications of the coalescence in the system obtained by changing the carrier fluid property: indeed, local large and low shear rates are found in the regions between two interacting droplets for shear-thickening and thinning fluids, respectively, resulting in increased and reduced local viscosity which ultimately affects the drainage time of the system. This process is independent of the nominal viscosity ratio of the two fluids and we show that it can not be understood by considering only the mean shear rate and viscosity of the two fluids across the domain, but the full spectrum of shear rate must be taken into account.
\end{abstract}

\maketitle

\section{Introduction} \label{sec:introduction}
Emulsions are mixtures of two or more liquids that are partially or totally immiscible. They are present in many biological and industrial applications such as waste treatment, oil recovery and pharmaceutical manufacturing and are also relevant in the field of colloidal science where the accuracy and the control of the production process of functional materials rely on the knowledge of the complex microstructure of the suspension \cite{xia_gates_yin_lu_2000a}. In this work we focus on the rheology of biphasic emulsions by means of direct numerical simulations. \mr{The study of two-phase flows has recently become of utmost importance related to the ongoing COVID- 19 pandemic, caused by airborne transmission of virus-containing saliva droplets transported by human exhalations \cite{dbouk_drikakis_2020q,  das_alam_plumari_greco_2020a, rosti_olivieri_cavaiola_seminara_mazzino_2020l, rosti_cavaiola_olivieri_seminara_mazzino_2021n}}

The study of rheology is motivated by the many fluids in nature and industrial applications which exhibit a non-Newtonian behavior, \ie a nonlinear relation between the shear stress and the shear rate; the relation between these macroscopic behaviors and the microstructure is often studied assuming suspensions of objects in a Newtonian solvent with dynamic viscosity $\mu$. \citet{einstein_1911a} was the first to provide a closure for the effective viscosity $\mu_e$ of a dilute rigid particle suspensions with vanishing inertia, and showed theoretically that $\mu_e$ is a linear function of the particle volume fraction $\Phi$ 
\begin{equation} \label{eq:effvis_einstein}
  \mu_e = \mu \left( 1 + \frac{5}{2} \Phi \right).
\end{equation}
Non-rigid and deformable objects, such as deformable particles \cite{rosti_brandt_2018a}, capsules \cite{matsunaga_imai_yamaguchi_ishikawa_2016a} and droplets \cite{rosti_de-vita_brandt_2019a} behave differently because of their deformation and in the latter case also coalescence and breakup. \citet{taylor_1932a} extended \equref{eq:effvis_einstein} by introducing the viscosity ratio between the two phases $\lambda$ (defined as the disperse phase over the matrix phase viscosity), thus obtaining
\mr{\begin{equation} \label{eq:effvis_taylor}
  \mu_e = \mu \left( 1 + \frac{5}{2} \Phi \frac{\lambda+\frac{2}{5}}{\lambda+1} \right),
\end{equation}}
which for a unit viscosity ratio reduces to
\mr{\begin{equation} \label{eq:effvis_taylor2}
  \mu_e = \mu \left( 1 + \frac{7}{4} \Phi \right).
\end{equation}}
\citet{pal_2002a, pal_2003a} derived expressions to evaluate the effective viscosity of infinitely dilute and concentrated emulsions using the effective medium approach. These relations assume limiting cases to model surface tensions effects (either going to zero or infinite) and thus direct numerical simulation is a valuable tool to overcome these limitations.

\citet{zhou_pozrikidis_1993a} simulated numerically two-dimensional emulsions and their work was later extended to three-dimensional flows by \citet{loewenberg_hinch_1996a}. Their results revealed the complexity of emulsions rheology, with pronounced shear thinning and large normal stresses associated with an anisotropic microstructure resulting from the alignment of deformed drops in the flow direction. More recently, \citet{srivastava_malipeddi_sarkar_2016a} also investigated the inertial effects on emulsions and found that, while in the absence of inertia emulsions display positive first and negative second normal stress differences, small amount of inertia alters their signs with the first normal stress difference becoming negative and the second one positive. The sign change is caused by the increasing drop inclination in the presence of inertia, which in turn directly affects the interfacial stresses. These previous studies were focused only on deforming droplets, without taking into account coalescence and breakup, which however plays a key role for the rheological behaviour of emulsions and are the main elements distinguishing emulsions from particle suspensions.

The effect of coalescence was tackled numerically more recently by \citet{rosti_de-vita_brandt_2019a} and \citet{de-vita_rosti_caserta_brandt_2019a} who showed that the effective viscosity $\mu_e$ of a droplet suspension can be properly described by extending \equref{eq:effvis_einstein} to the second order similarly to what proposed by \citet{batchelor_green_1972a} for dilute particle suspensions, \ie
\mr{\begin{equation} \label{eq:effvis_rosti}
  \mu_e = \mu_0 \left( 1 + c_1 \Phi + c_2  {\Phi}^2 \right),
\end{equation}}
where $c_1$ and $c_2$ are fitting parameters depending on the viscosty ratio $\lambda$ and capillary number $Ca$. While $c_2$ is positive in the case of rigid particles, indicating that the effective viscosity always grows with the volume fraction ($c_1$ is positive  consistently with \equref{eq:effvis_einstein} and \equref{eq:effvis_taylor}), on the other hand $c_2$ is negative for droplets, and thus $\mu_e$ is a concave function of $\Phi$, \ie $\mu_e$ has a maximum for an intermediate value of $\Phi$ and then decreases with the volume fraction. The volume fraction for which the maximum is reached decreases with the viscosity ratio $\lambda$ and $\mu_e$ is again a convex function of $\Phi$ in the case of droplets when the coalescence of the solute phase is suppressed \cite{de-vita_rosti_caserta_brandt_2019a}; in this case, their behaviour resembles what found for deformable particles \cite{rosti_brandt_mitra_2018a}, thus confirming that coalescence is a key mechanism that can dramatically change the rheology of emulsions. \mr{The role of coalescence in a turbulent channel flow has been recently investigated numerically by Cannon \etal  \cite{ianto-cannon_rosti_2021u}.}

In this works we account for the coalescence and study how it affects the rheology of emulsions adding an additional complexity to the system by considering non-Newtonian solvents \mr{\cite{mohammad-karim_2020c, suo_jia_2020u}.} Few theoretical expressions were proposed in the past to predict the rheology of viscoelstic emulsions \cite{kerner_1956a, palierne_1990a, bousmina_1999a}. \citet{palierne_1990a} derived an expression to predict the shear modulus of emulsions of viscoelastic materials accounting for the mechanical interactions between inclusions. A different expression which however provides similar predictions was later derived by \citet{bousmina_1999a} who extended Kerner's model \cite{kerner_1956a} for flow of composite elastic media to emulsions of viscoelastic phases with interfacial tension undergoing small deformations. This model is able to predict some general features typical of viscoelastic emulsions, such as shoulders in the storage modulus $G'$ at low frequency and a long relaxation time process. However, emulsions with non-Newtonian fluids are characterised by several peculiar behaviours found experimentally which are usually difficult to capture with theoretical models; for example, \citet{mason_bibette_1996a} were able to produce monodisperse emulsions using a viscoelastic medium under shear; the authors showed the possibility of controlling the final droplet size by altering the shearing conditions and the emulsion viscoelasticity, suggesting that the monodispersity resulted from the suppression of the capillary instability until the droplet is sufficiently elongated. A similar study was later performed by \citet{zhao_goveas_2001a} who investigated the deformation and breakup of a dilute emulsion with a viscoelastic continuous phase under shear flow. Also these authors found a size selection mechanism of the resulting ruptured drops, not found in Newtonian fluids. To be best of our knowledge, no systematic study of the rheology of concentrated emulsions are available in the literature, with only very few studies focusing on rigid particles suspensions in viscoelastic media \cite{patankar_hu_2001a, zarraga_hill_leighton-jr_2001a, scirocco_vermant_mewis_2005a, davino_greco_hulsen_maffettone_2013a, yang_shaqfeh_2018a, alghalibi_lashgari_brandt_hormozi_2018a, rosti_brandt_2020a}, and the present work is aimed to fill in this gap.

The paper is structured as follow: in \secref{sec:formulation} we describe the governing equations and the numerical method used to numerically solve them. After discussing the chosen setup, in \secref{sec:result} we present the results of our simulations and discuss the effect of shear-thinning and thickening on the rheology of emulsions. Finally, we summarize the main findings and draw the main conclusions in \secref{sec:conclusion}.

\section{Methodology} \label{sec:formulation}
\begin{figure*}
  \centering
  \includegraphics[width=0.45\textwidth]{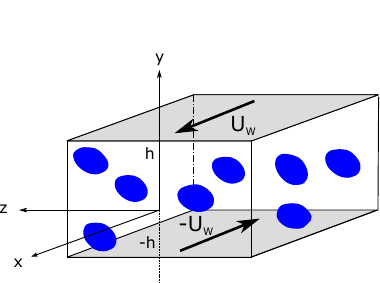}\\
  \includegraphics[width=0.95\textwidth]{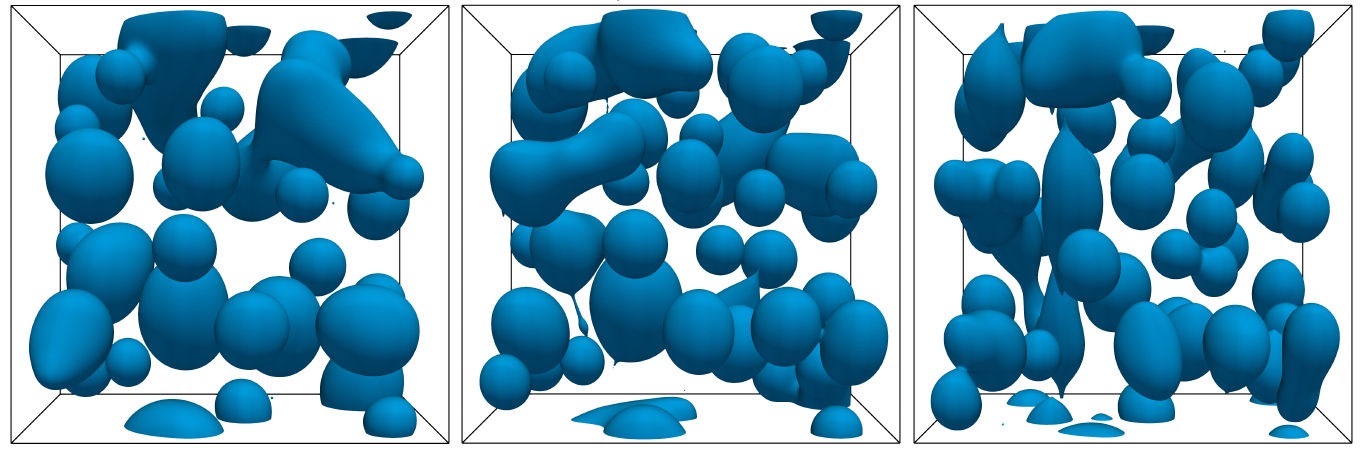}
  \caption{(Top) Sketch of the computational domain. (bottom) Top view ($x$-$z$ plane) of the computational domain with droplets shown in blue and the flow in the vertical direction. The left, middle and right panels show the instantaneous configuration of the droplets suspended in a shear-thinning $n=0.5$, Newtonian $n=1$ and shear-thickening $n=1.5$ fluid, respectively, at $t \approx 25 \dot{\gamma}_0^{-1}$ .}
  \label{fig:sketch}
\end{figure*}
We consider a flow with two incompressible viscous fluids separated by an interface in a channel with moving walls, \ie in a plane Couette geometry. \figrefC[top]{fig:sketch} shows a view of the geometry considered in the present analysis, where $x$, $y$ and $z$ ($x_1$, $x_2$, and $x_3$) denote the streamwise, wall-normal, and spanwise coordinates, while $u$, $v$ and $w$ ($u_1$, $u_2$, and $u_3$) denote the corresponding components of the velocity vector field. The lower and upper impermeable moving walls are located at $y=-h$ and $y=h$, respectively, and move in opposite direction with constant streamwise velocity $u = \pm U_w$, providing a nominal shear-rate $\dot{\gamma}_0 = 2U_w/2h$.

To identify the two fluid phases, we introduce an indicator (or color) function $H$. In particular, $H=1$ in the regions occupied by the fluid $1$ and $H=0$ otherwise. In this work, we assume fluid $1$ to be the solute (dispersed phase) and fluid $2$ the solvent (carrier phase). Since the interface is transported by the fluid velocity, we advect the indicator function as follows
\begin{equation} \label{eq:indicator_advection}
  \frac{\partial \phi}{\partial t}+\frac{\partial u_i H}{\partial x_i}=\phi \frac{\partial u_i}{\partial x_i},
\end{equation}
where $\phi$ is the cell-averaged value of $H$ also called the volume of fluid function.  The two fluids are governed by the momentum conservation and the incompressibility constraint and which can be written as one set of equations incorporating the interface conditions, which read 
\begin{equation} \label{eq:navier_stokes}
  \rho \left( \frac{\partial u_i}{\partial t} + \frac{\partial u_i u_j}{\partial x_j} \right) = \frac{\partial \sigma_{ij}}{\partial x_j} + f_i \;\;\;\; \frac{\partial u_i}{\partial x_i} = 0,
\end{equation}
Here, $\rho$ is the fluid density assumed to be the same in the two phases, $\sigma_{ij}$ the Cauchy stress tensors and $f_i = \sigma \kappa n_i \delta \approx \sigma \kappa \partial \phi / \partial x_i$ is added to account for the interface condition \citep{brackbill_kothe_zemach_1992a}, where $\sigma$ is the interfacial surface tension coefficient, $\kappa$ and $n_i$ the local curvature and unit normal vector of the interface and $\delta$ is the Dirac's delta function at the interface. The Cauchy stress tensor $\sigma_{ij}$ is defined as \mr{$\sigma_{ij} = -p \delta_{ij} + 2 \mu_l \mathcal{D}_{ij}$} where $p$ is the pressure, $\mathcal{D}_{ij}=\left(\partial u_i/\partial x_j+\partial u_j\partial x_i\right)/2$ the strain rate tensor and \mr{$\mu_l=\phi \mu_1 + \left( 1 - \phi \right) \mu_2$ the local dynamic viscosity} written in a mixture form as a simple weighted average based on the local value of $\phi$ of the dynamic viscosity in the two phases $\mu_1$ and $\mu_2$. The solute fluid $1$ is assumed to be Newtonian, and its dynamic viscosity $\mu_1$ is a constant, while the solvent fluid $2$ is a non-Newtonian fluid described by a power law model where the local value of $\mu_2$ is a function of the local shear-rate $\dot{\gamma}_l$.

The equations of motion are solved on an uniform staggered fixed Eulerian grid where the fluid velocity components are located on the cell faces and all other variables (density, viscosity, pressure, stress and volume of fluid) at the cell centers. The time integration is performed with a fractional-step method based on the second-order Adams-Bashforth scheme. The indicator function $H$ needed to capture the interface between the two fluids is found following the multidimensional tangent of hyperbola for interface capturing (MTHINC) method proposed by \citet{ii_sugiyama_takeuchi_takagi_matsumoto_xiao_2012a} and transported with the volume of fluid method. More details on the numerical implementation can be found in \citet{rosti_de-vita_brandt_2019a}.

\subsection{Non-Newtonian fluid model}
Fluids that exhibit a non-linear behavior between the shear stress and the shear rate are called non-Newtonian and, in particular, fluids whose response does not depend explicitly on time but only on the present shear rate are denoted generalized Newtonian fluids. When the shear stress increases more than linearly with the shear rate, the fluid is called dilatant or shear-thickening, whereas in the case of opposite behavior, \ie when the shear stress increases less than linearly with the shear rate, the fluid is called pseudoplastic or shear-thinning. Several models have been developed to capture the different behaviours of various non-Newtonian fluids and in the current study, we focus on the simple inelastic power law models, where the local viscosity of the fluid is a function of the sole local value of shear rate. A relation that can be used to summarize the behaviors previously described for complex fluids is
\begin{equation} \label{eq:powerlaw}
\mu_{pl} = K \dot{\gamma_l}^{n-1},
\end{equation}
where $n$ is the flow index and $K$ the fluid consistency index. A Newtonian behavior is recovered when $n=1$ and $K=\mu$, while values of the flow index above and below unity, $n>1$ and $n<1$, denote shear-thickening and shear-thinning fluids, respectively. The consistency index $K$ measures how strong the fluid responds to the imposed deformation rate but it is not possible to compare different values of $K$ for fluids with different flow indexes $n$ because its dimension is a function of $n$ itself. In general, the local viscosity of the non-Newtonian fluid increases with $\dot{\gamma}_l$ for shear-thickening fluids, while it reduces for shear-thinning ones, which means that the fluidity of shear-thickening fluids reduces increasing the shear rate, while the opposite is true for shear-thinning fluids. In the previous relation, the local shear rate $\dot{\gamma}_l$ is the second invariant of the strain-rate tensor $\mathcal{D}_{ij}$ and is computed by its dyadic product, \ie $\dot{\gamma}_l=\sqrt{2 \mathcal{D}_{ij} \mathcal{D}_{ij}}$ \cite{bird_armstrong_hassager_1987a}. The viscosity of a power law shear-thinning fluid becomes infinite for null shear-rate; to overcome this numerical issue, the Carreau fluid  model is used instead in which the local viscosity is computed as
\begin{equation}
\mu_{ca} = \mu_\infty + \left( \mu_0 - \mu_\infty \right) \left[ 1+ \left( k \dot{\gamma}_l \right)^2 \right]^{(n-1)/2}.
\label{eq:carreau}  
\end{equation}
In this equation, $\mu_\infty$ and $\mu_0$ indicate the lower and upper limits of fluid viscosity at infinite and zero shear rates. The flow index $n$ characterizes the behaviour of the fluid: for $n<1$ the fluid is shear thinning and the material time constant $k$ represents the degree of shear-thinning. More details on the Carreau and power-low models can be found in \cite{morrison_2001a}.

\subsection{Setup}
\begin{figure}
  \centering
  \includegraphics[width=0.45\textwidth]{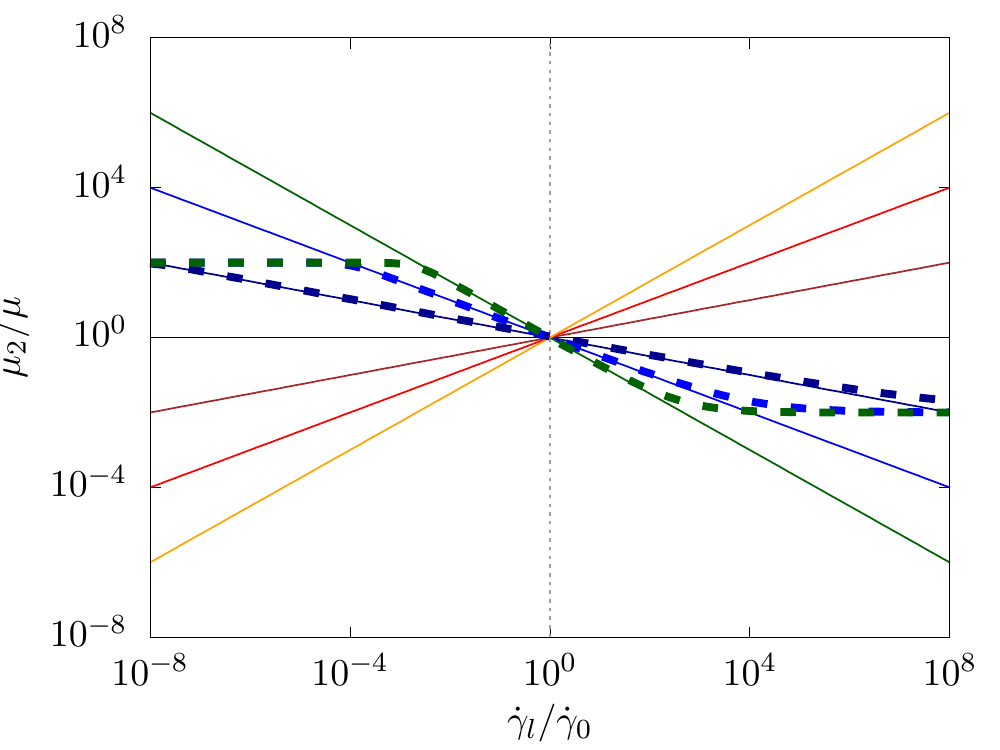}
  \caption{Local viscosity of the carrier fluid $\mu_2$ normalised by the reference Newtonian value $\overline{\mu}_2$ as a function of the \mr{local shear rate $\dot{\gamma}_l$} normalised by the reference value $\dot{\gamma}_0$. The green, blue, dark-blue, black, brown, red and orange colors are used to distinguish fluids with different flow index $n$ equal to $0.25$, $0.5$, $0.75$, $1$, $1.25$, $1.5$ and $1.75$, respectively, while the solid and dashed lines are used to distinguish power law and Carreau fluids.}
  \label{fig:shear0}
\end{figure}
We consider a two-phase system in a Couette flow consisting of a suspension of droplets, see \figrefS{fig:sketch}. The droplets are initially spherical with radius $\mathcal{R}$ and are randomly distributed in the computational domain, which is a rectangular box of size $16 \mathcal{R} \times 10 \mathcal{R} \times 16 \mathcal{R}$ in the stream, wall-normal and spanwise directions. The computational domain is discretized with a uniform Cartesian grid with $16$ grid points per initial droplet radius $\mathcal{R}$ (\ie $160$ grid points per channel height $2h$).  Note that,  the actual resolution of the droplets dynamic improves as droplets coalescence and grow in size. The suspended fluid is Newtonian with uniform constant viscosity $\mu_1$, while three different kinds of carrier fluids are studied: Newtonian fluids with viscosity $\mu_2=\overline{\mu}$, shear-thickening fluids with viscosity $\mu_2=\mu_{pl}$ defined by \equref{eq:powerlaw} and shear-thinning fluids with viscosity $\mu_2=\mu_{ca}$ defined by \equref{eq:carreau}. The flow index $n$ is varied in the range $0.25 \div 1.75$ and we choose to define the three fluids such that the viscosity at the applied reference shear rate $\dot{\gamma}_0$ is the same in all of them, \ie $\overline{\mu}$. Thus, the parameter $K$ in the power law fluid is determined by imposing $\overline{\mu} = \mu_{pl} \left( \dot{\gamma}_0 \right)$ (yielding $K=\overline{\mu} \dot{\gamma}_0^{1-n}$), and similarly the Carreau fluid model parameter $k$ is found by imposing $\overline{\mu} = \mu_{ca} \left( \dot{\gamma}_0 \right)$. Note that, in our simulations with the Carreau model we have fixed the ratio $\mu_0/\mu_\infty$ to $10^4$. \figrefSC{fig:shear0} shows the rheology of all the carrier fluids considered in the present study. We vary the volume fractions of the solute phase $\Phi$ (defined as $\langle \phi \rangle_{xyz}$ with the brackets indicating the average operator) in the range $0 \div 0.3$ and we consider two different viscosity ratios $\lambda=\mu_1 / \overline{\mu}$ equal to $1$ and $0.01$. The surface tension coefficient $\sigma$ is fixed, such that the droplets have an initial capillary numbers $Ca=\overline{\mu} \dot{\gamma}_0 \mathcal{R} / \sigma$ equal to $0.2$. Finally, the Reynolds number $Re= \rho \dot{\gamma}_0 \mathcal{R}^2 / \overline{\mu}$ is fixed to $0.1$, so that we can consider the inertial effects negligible.

Topological changes of droplets such as coalescence and breakup are treated here naturally without any additional model, since with the volume of fluid method these happen naturally when the distance between two interface falls within a grid cell.  However, this means that the coalescence and breakup processes can be strongly influenced by the grid size resolution; a detailed discussion of this matter is reported in the appendix. Note also that, the parameters are chosen similar to that of previous works that can be found in the literature to ease comparisons \citep{rosti_brandt_mitra_2018a, rosti_brandt_2018a, rosti_de-vita_brandt_2019a, de-vita_rosti_caserta_brandt_2019a}. Finally, the numerical code used in the present work has been tested and validated in the past in several works for laminar and turbulent flows of single and multiphase systems \citep{izbassarov_rosti_niazi-ardekani_sarabian_hormozi_brandt_tammisola_2018a, rosti_brandt_mitra_2018a, rosti_ardekani_brandt_2019a, rosti_brandt_2020a, chiara_rosti_picano_brandt_2020a} \footnote{Several validations can be also found at https://groups.oist.jp/cffu/validation}.

\section{Results} \label{sec:result}
\begin{figure}
  \centering
  \includegraphics[width=0.45\textwidth]{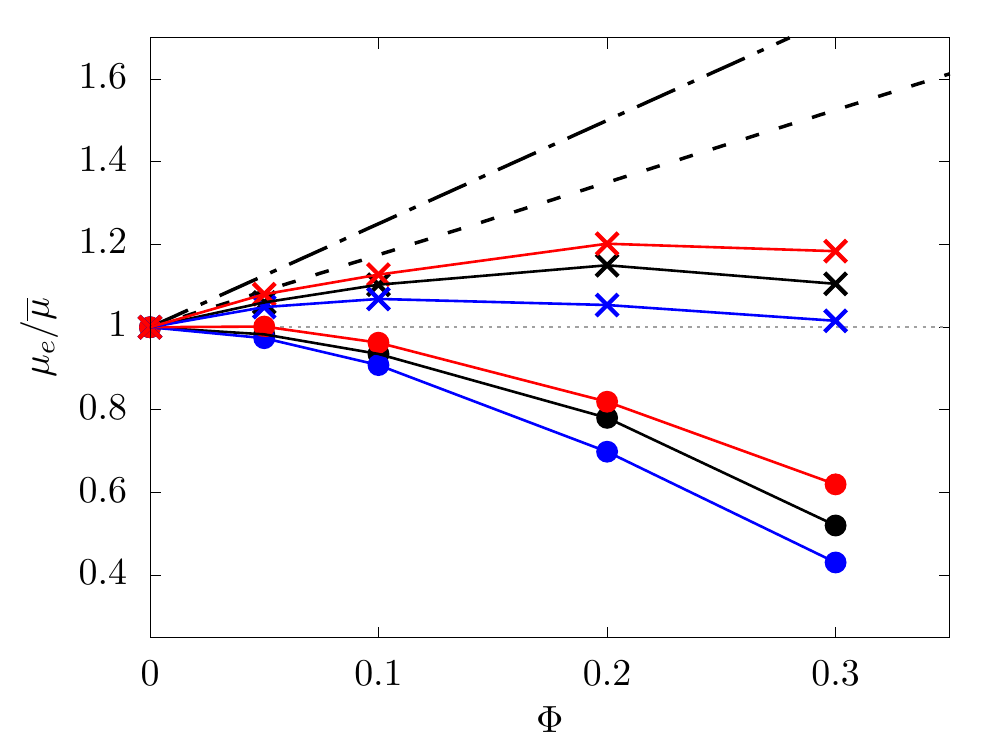}
  \caption{Normalised effective viscosity $\mu_e$ as a function of the droplets volume fraction $\Phi$ for different fluids and viscosity ratios. The color scheme is the same used in \figrefS{fig:shear0}, with the blue, black and red colors indicating $n=0.5$, $1$ and $1.5$, while the cross $\pmb \times$ and circle $\bullet$ symbols are used to distinguish $\lambda$ equal to $1$ and $0.01$, respectively.  The black dash-dotted and dashed lines are Einstein and Taylor equations \ref{eq:effvis_einstein} and \ref{eq:effvis_taylor2}.}
  \label{fig:visc}
\end{figure}
We study the rheology of a droplet suspension immersed in power law fluids and compare the results with those obtained in a Newtonian fluid. To do so, we start our rheological analysis by considering the effective viscosity $\mu_e$ which is the viscosity of a Newtonian fluid that gives the same shear stress at the nominal shear rate, and is thus defined as
\begin{equation} \label{eq:effective_vis}
  \mu_e = \frac{\langle \sigma_{12}^w \rangle_{xz,t}}{\dot{\gamma}_0},
\end{equation}
where $\sigma_{12}^w$ is the shear component of the stress tensor evaluated at the walls. In general, we expect the rheology of a two fluid system to be a function of the Reynolds number $Re$, the capillary number $Ca$, the viscosity ratio $\lambda$, the solute volume fraction $\Phi$, the confinement ratio $2\mathcal{R}/2h$ and the non-Newtonian property of the carrier fluid here summarised by the power law index $n$. In the present work, we limit our analysis to inertialess flows, \ie $Re \lll 1$, the capillary number $Ca$ is not varied and fixed to a value such that a single droplet of size $\mathcal{R}$ subject to the applied shear rate $\dot{\gamma}_0$ does not breakup,  and the domain size is chosen such that confinement effects are negligible. Thus, we can simplify the analysis to $\mu_e \approx \mathcal{F} \left( \Phi, \lambda, n \right)$.

\figrefSC{fig:visc} shows the effective viscosity $\mu_e$ as a function of the droplet volume fraction $\Phi$ for two different viscosity ratio $\lambda$ and for three different power law index $n$. In the Newtonian carrier fluid (black color), the behaviour is the same observed by \citet{de-vita_rosti_caserta_brandt_2019a}: when the viscosity of the two fluids is the same, \ie $\lambda=1$, the effective viscosity $\mu_e$ first grows with the volume fraction similarly to a rigid particle suspension, then reaches a maximum value for an intermediate $\Phi$ and then starts decreasing again. When the viscosity of the dispersed phase is reduced, \ie $\lambda$ decreases, the volume fraction for which the maximum effective viscosity is reached reduces, and for a sufficiently low $\lambda$ the maximum is reached at $\Phi \approx 0$ and the effective viscosity curve decreases with $\Phi$. Note that, the effective viscosity $\mu_e$ can be even smaller than the carrier fluid one $\overline{\mu}$ in the case of $\lambda<1$. The non-monotonic behaviour of $\mu_e$ with $\Phi$ is a direct consequence of the limiting behavior for $\Phi \rightarrow 0$ and $\Phi \rightarrow 1$ where $\mu_e$ is equal to $\mu_2$ and $\mu_1$ by definition \cite{rosti_de-vita_brandt_2019a}. When the carrier fluid is non-Newtonian the situation is further modified. In general, for all the power law index $n$ and viscosity ratio $\lambda$ considered in the present study, we observe that the non-monotonic behaviour is preserved and that the effective viscosity $\mu_e$ is larger for shear-thickening fluids with $n>1$ and smaller for shear-thinning fluids with $n<1$ than their Newtonian counterparts with $n=1$. The difference in $\mu_e$ between the Newtonian and non-Newtonian fluids grows with the volume fraction $\Phi$ and larger difference are evident between the Newtonian and shear-thinning fluids than what observed with the shear-thickening fluid for both the viscosity ratio $\lambda$ considered. These changes of $\mu_e$ with $n$ are qualitatively similiar to what observed for a rigid particle suspension immersed in a power law carrier fluid \cite{alghalibi_lashgari_brandt_hormozi_2018a}, except for the non-monotonicity of the viscosity curve which is due to the coalescence mechanism in the emulsions and absent in particle suspensions.

\begin{figure}
  \centering
  \includegraphics[width=0.45\textwidth]{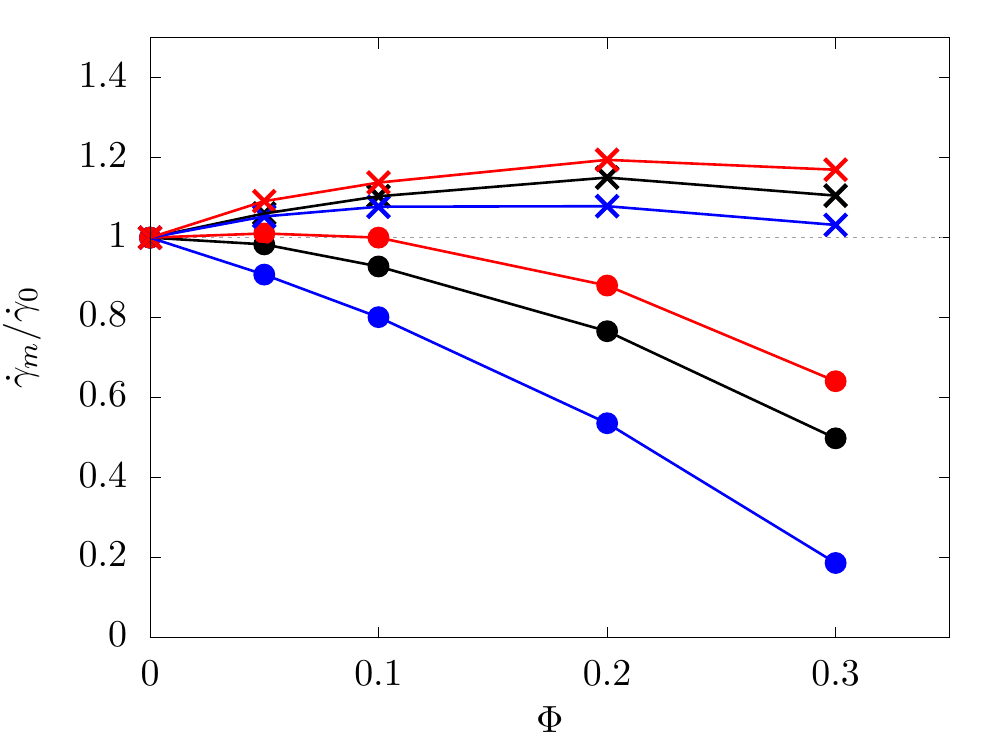}
  \includegraphics[width=0.45\textwidth]{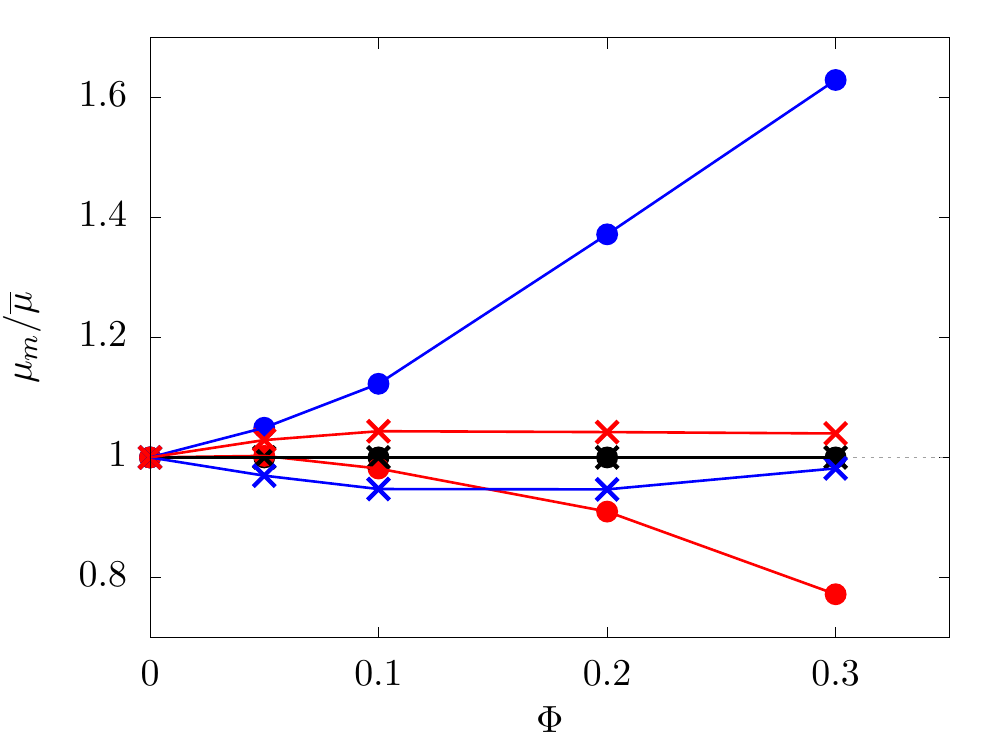}
  \caption{Normalised \mr{mean (top) shear rate $\dot{\gamma}_m$ and (bottom) viscosity $\mu_m$} as a function of the droplet volume fraction $\Phi$. The color and symbol schemes are the same used in \figrefS{fig:visc}, with the blue, black and red colors indicating $n=0.5$, $1$ and $1.5$, while the cross $\pmb \times$ and circle $\bullet$ symbols indicating $\lambda=1$ and $0.01$.}
  \label{fig:local}
\end{figure}
To understand the change in the effective viscosity curve with $n$, we compute the \mr{mean shear rate $\dot{\gamma}_m = \langle \phi \dot{\gamma}_l \rangle_{xyz,t} / \langle \phi \rangle_{xyz,t}$ and viscosity $\mu_m = \langle \phi \mu_l \rangle_{xyz,t} / \langle \phi \rangle_{xyz,t}$ in the carrier phase} which are shown in \figrefS{fig:local}. In the figure we observe that the \mr{mean shear rate} exhibits a behaviour similar to the effective viscosity $\mu_e$: \mr{$\dot{\gamma}_m$} grows with the volume fraction $\Phi$, then reaches a maximum value and finally decreases for large $\Phi$ for $\lambda = 1$, while for $\lambda <1$ the maximum is reached at lower volume fraction and for $\lambda=0.01$ the \mr{mean shear rate} deceases for all $\Phi$. Also the effect of the flow index is analogous to what previously observed for $\mu_e$, and indeed \mr{$\dot{\gamma}_m$} is larger for $n>1$ and smaller for $n<1$ than the Newtonian case with $n=1$, with larger differences evident for the shear-thinning fluids. Less trivial is the behaviour of the \mr{mean viscosity $\mu_m$}; for $\lambda \ge 1$, being the shear rate larger than the reference one $\dot{\gamma}_0$, the \mr{mean viscosity} in the shear-thickening fluid is larger than the Newtonian one while it is smaller in the shear-thinning fluid, as expected from the rheology of the fluid previously reported in \figrefS{fig:shear0}. On the other hand, when $\lambda < 1$, the \mr{mean shear rate} is actually smaller than $\dot{\gamma}_0$ and thus the behaviour of the non-Newtonian fluids is opposite, with the \mr{mean viscosity} of the shear-thinning fluid being larger than the Newtonian one and smaller for the shear-thickening fluid. Thus, while for $\lambda \ge 1$ the \mr{mean viscosity and shear rate} are both contributing to increase or decrease the effective viscosity of the suspension, when $\lambda < 1$ because of this peculiar behaviour, \mr{the mean viscosity} is actually counteracting and reducing the effect of the change in \mr{mean shear rate}. This reverse trend obviously never happens in a particle suspension, where the shear rate is always larger than the single-phase value in the presence of suspended particles, and is thus a prerogative of emulsions. 

\begin{figure}
  \centering
  \includegraphics[width=0.45\textwidth]{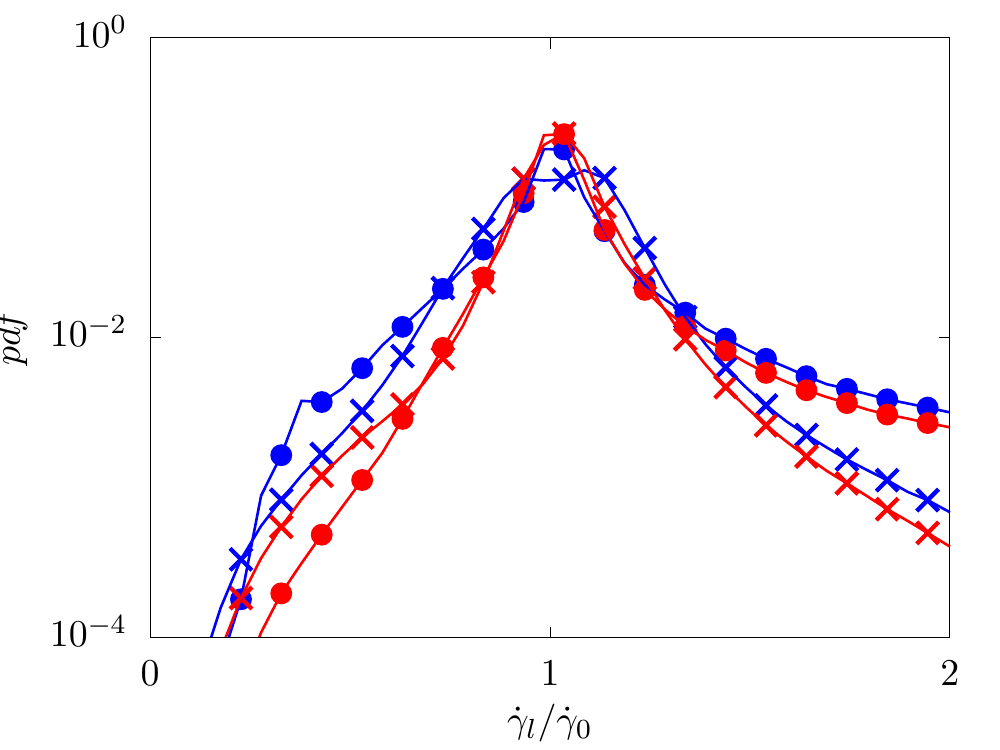}
  \includegraphics[width=0.45\textwidth]{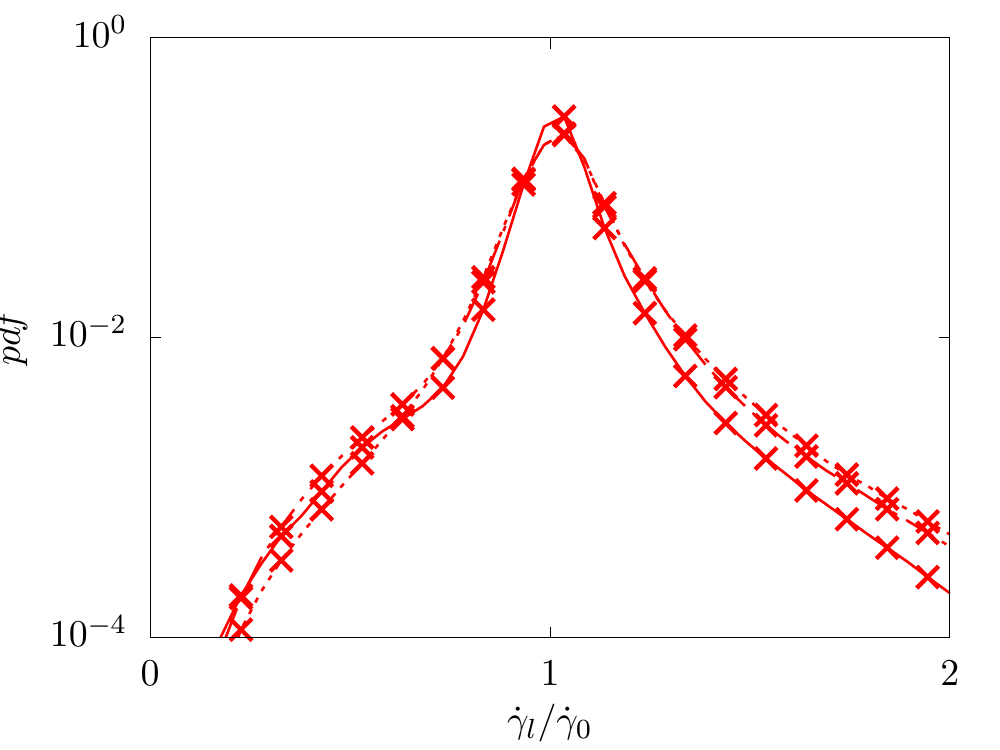}
  \caption{(top) The probability density function $pdf$ of the \mr{local shear rate $\dot{\gamma}_l$ in the carrier phase} for different carrier fluids at $\Phi=10\%$. (bottom) The probability density function $pdf$ of the shear rate \mr{$\dot{\gamma}_l$ in the carrier phase} for different volume fractions $\Phi$ in a shear-thickening fluid with $\lambda=1$. The color and symbol schemes in the figures are the same used in \figrefS{fig:visc}, with the blue, black and red colors indicating $n=0.5$, $1$ and $1.5$, while the cross $\pmb \times$ and circle $\bullet$ symbols indicating $\lambda=1$ and $0.01$. Also, the solid, dashed and dotted lines in the bottom figure are used to distinguish $\Phi=5\%$, $10\%$ and $20\%$.}
  \label{fig:pdf}
\end{figure}
Notwithstanding the counter-effect of the \mr{mean viscosity} for $\lambda < 1$, the effective viscosity $\mu_e$ in the shear-thinning and thickening fluids is always smaller and larger than in the Newtonian fluid, respectively. In order to understand what is really affecting the emulsion viscosity, we need to study the value of the local shear rate not in terms of its mean value but by considering its full range of values assumed. \figrefC[top]{fig:pdf} reports the probability density function $pdf$ of the shear rate $\dot{\gamma}_l$ \mr{in the carrier phase} for all the fluids studied with different $n$ and $\lambda$ and a selected volume fraction $\Phi$ equal to $10\%$. The graphs show that a large spectrum of shear rate exists for all the fluids, with the $pdf$ characterised by a strong peak at its mean value (see \figref[top]{fig:local}) around the nominal shear rate $\dot{\gamma}_0$ and with a positive skewness, \ie with more high shear rates than low ones. The latter result indicates an higher probability of finding low local viscosity in a shear-thinning fluid and of high viscosity in a shear-thickening one, independently of the viscosity ratio considered.
\begin{figure*}
  \centering
  \includegraphics[width=0.95\textwidth]{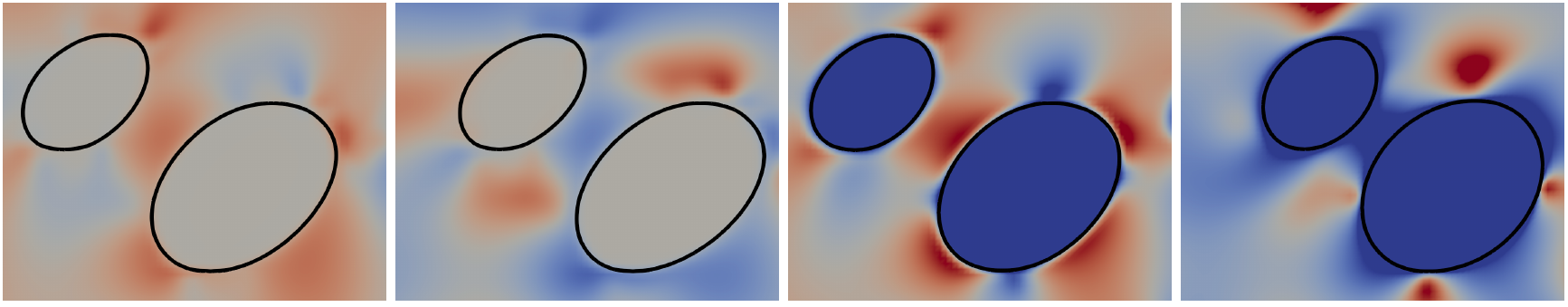}
  \caption{Instantaneous side view visualisation of the droplets (solid black line) and color contour of the local viscosity in the two phases \mr{$\mu_l/\overline{\mu}$}. The color scale goes from $0.6$ to $1.4$; the first two panels correspond to $\lambda=1$ and the last two to $\lambda=0.01$. For each set of $\lambda$, the first panel is the shear-thickening fluid case and the second one the shear-thinning fluid case.}
  \label{fig:slice}
\end{figure*}
The origin of these high shear rate events is the droplet interaction. To prove this, we show in \figrefS{fig:slice} some instantaneous visualisation of the local viscosity in the solvent and solute phases. As we can clearly see from the figure, when two droplets interact in a shear-thickening fluid (panel $1$ and $3$) the local viscosity in the interstitial region between the two is large independently of the value of viscosity ratio $\lambda$ considered, thus indicating a local high shear rate, while on the contrary when the two interacting droplets are suspended in a shear-thinning fluid (panel $2$ and $4$) the local viscosity in between is low, thus indicating a local low shear rate. Since the occurrence of the high shear rate events is linked to the droplet interactions, it is a function of the volume fraction $\Phi$ and grows with it, as shown in \figref[bottom]{fig:pdf}.

\begin{figure}
  \centering
  \includegraphics[width=0.45\textwidth]{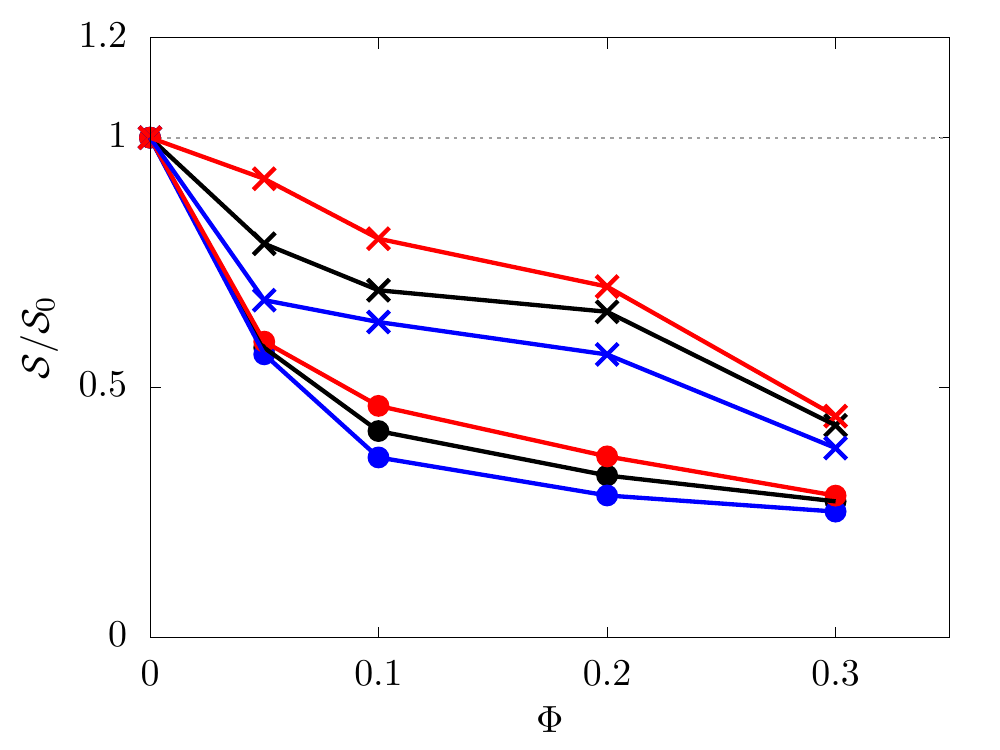}
  \includegraphics[width=0.45\textwidth]{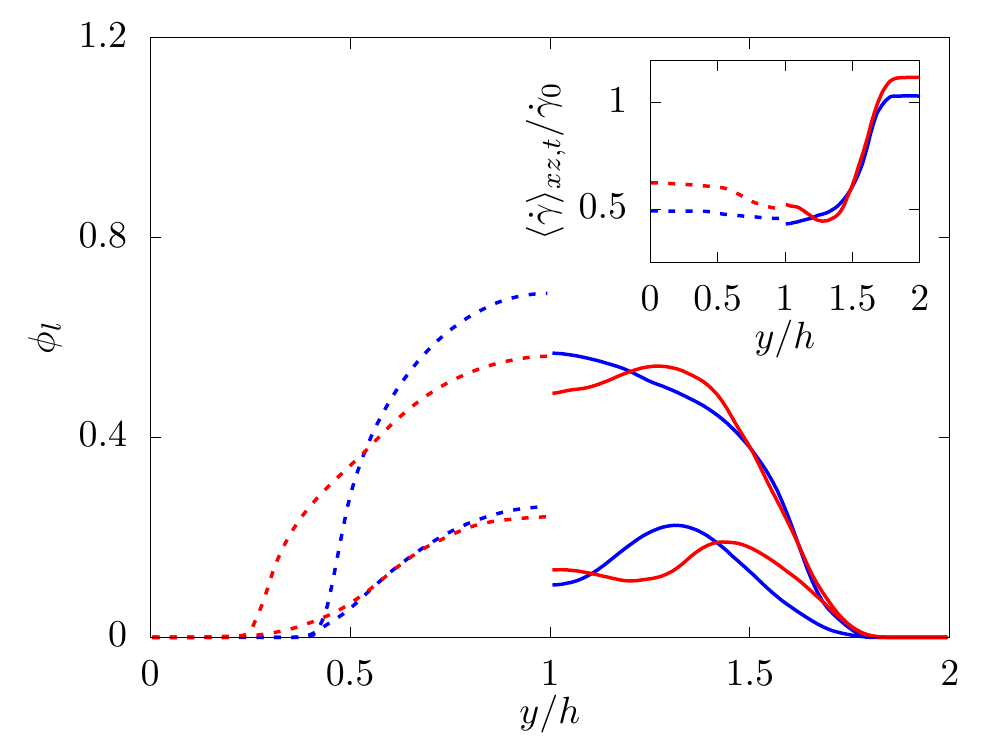}
  \caption{(top) The total surface area $\mathcal{S}$ normalised with its initial value $\mathcal{S}_0$ as a function of the droplet volume fraction $\Phi$. The color and symbol schemes in the figures are the same used in \figrefS{fig:visc}, with the blue, black and red colors indicating $n=0.5$, $1$ and $1.5$, while the cross $\pmb \times$ and circle $\bullet$ symbols indicating $\lambda=1$ and $0.01$. (bottom) Averaged wall-normal concentration profile $\phi_l=\langle \phi \rangle_{xz,t}$. The red and blue colors distinguish the cases with the shear-thickening $n=1.5$ and shear-thinning $n=0.5$ carrier fluid, and the solid and dashed lines the two viscosity ratios $\lambda=1$ and $0.01$. Two representative volume fraction are considered, $\Phi=0.1$ and $0.3$. The inset figure shows the wall-normal distribution of the \mr{mean shear rate} for the case with highest volume fraction, $\phi=0.3$.}
  \label{fig:surf}
\end{figure}
The different viscosity found for shear-thickening and thinning fluids in the interstitial region between two interacting droplets affects the coalescence mechanism. \citet{chesters_1991a} was the first to develop a theoretical framework to study the interaction of droplets. He proposed that the dynamics of two colliding droplets is an interplay of an external flow - the driving flow - responsible of the collision, and an internal flow - the drainage film between the two droplets - responsible of the interface deformation and rupture. The former is described in terms of a collision duration $\tau_c$, roughly proportional to the inverse of the shear rate, \ie $\tau_c \dot{\gamma} \approx 1 $ \cite{smoluchowski_1918a}, the latter in terms of a drainage time $\tau_d$ which by scaling analysis can be shown to be proportional to the capillary number $Ca$ of the droplets, \ie $\tau_d \dot{\gamma} \approx Ca^m$ where $m = 4/3$ if the drainage film is assumed to be flat \cite{chesters_1991a} or $m = 1$ for a dimpled-film shape \cite{frostad_walter_leal_2013a}. When the collision duration is larger than the drainage time $\tau_c > \tau_d$, droplets coalesce whereas in the opposite case $\tau_c < \tau_d$ they repel. In the former condition the emulsion is often called attractive, while in the latter it is defined repulsive \cite{guido_simeone_1998a}. These ideas are based on several simplifying hypothesis, with the main ones being almost spherical droplets and head on collision; notwithstanding this, the theory proved able to estimate the general coalescence behaviour of emulsions and we will use it here as well. For all the viscosity ratio considered, the viscosity in the drainage film between two droplets is lower in the shear-thinning fluid and larger in the shear-thickening one than their Newtonian counterpart; because of this change, the capillary number $Ca$ relevant for the collision is smaller/larger in the shear-thinning/thickening fluid than in the Newtonian case with a consequent decrease/increase of the drainage time $\tau_d$. Thus, Chester's theory predicts an increase of the coalescence in shear-thinning fluids and a decrease in the shear-thickening ones. Note however that, if one considers only the mean value \mr{$\mu_m$} reported in \figrefS{fig:local}, than this results would be opposite for the cases with $\lambda < 1$. To disentangle this opposite result, we quantify the droplets coalescence in our simulations by measuring the mean total surface area $\mathcal{S}$ in all the considered fluids, reported in \figref[top]{fig:surf}. Note that, a reduction of surface area indicates droplet merging and thus a reduction of the number of droplets. As shown in the figure, the surface area decreases with the droplet volume fraction $\Phi$, thus indicating the merging of small droplets into large ones, and as expected is larger in the cases with unit viscosity ratio $\lambda =1$ than in those with $\lambda=0.01$ \cite{de-vita_rosti_caserta_brandt_2019a}. When considering different fluids, \ie different power law index $n$, we clearly observe that when the fluid is shear-thickening the surface area $\mathcal{S}$ is always larger than in the Newtonian case, while on the contrary when the fluid is shear-thinning $\mathcal{S}$ is smaller than in the Newtonian case. This result is consistent with the prediction based on Chester's theory when using the local viscosity in the region in between droplets, \ie the right tale of the $pdf$ of the shear rate in \figref[top]{fig:pdf}, while it does not agree when considering only the mean shear (viscosity) value in \figrefS{fig:local}. It is thus important to consider the full-spectrum of available shear rates in the domain when studying power-law fluids and theory and models uniquely based on mean values would result in wrong  predictions.

\figrefC[bottom]{fig:surf} shows how the droplets redistribute across the domain for the different fluids analysed. Two representative volume fractions are considered, a dilute case with $\Phi=0.1$ and the most concentrated case studied with $\Phi=0.3$. In general, we observe that droplets tend to concentrate in the middle of the domain away from the walls. The migration of a droplet away from the wall in a shear flow was predicted theoretically by \citet{magnaudet_takagi_legendre_2003a} who provided an analytical expression for the the migration velocity; in our case, coalescence further enhance the concentration at the channel centerline in agreement with previous numerical studies in the literature \cite{rosti_de-vita_brandt_2019a, de-vita_rosti_caserta_brandt_2019a, alghalibi_rosti_brandt_2019a, de-vita_rosti_caserta_brandt_2020a}. Indeed, the concentration in the center of the domain is larger when the viscosity ratio $\lambda$ is small, due to the tendency of droplets to merge and form large conglomerates, while this is reduced for large values of $\lambda$ where the coalescence process is slowed down. In the latter case, we also find a reminiscence of the formation of wall-parallel layers as typically found in rigid particle suspensions which tend to form particle layers. When changing the carrier fluid property, we note that for shear-thickening fluids the concentration is always less peaked than in the shear-thinning case, thus further confirming the reduced level of coalescence and a more rigid-like behaviour of the droplets. The presence of the suspended droplets and their distribution in the channel affect the shear-rate,  which differently from a single phase flow where it is constant,  shows a strong inhomogeneity across the wall-normal direction, as reported in the inset of \figref[bottom]{fig:surf}.

%\subsection{The effect of the flow index $n$}
\begin{figure*}
  \centering
  \includegraphics[width=0.45\textwidth]{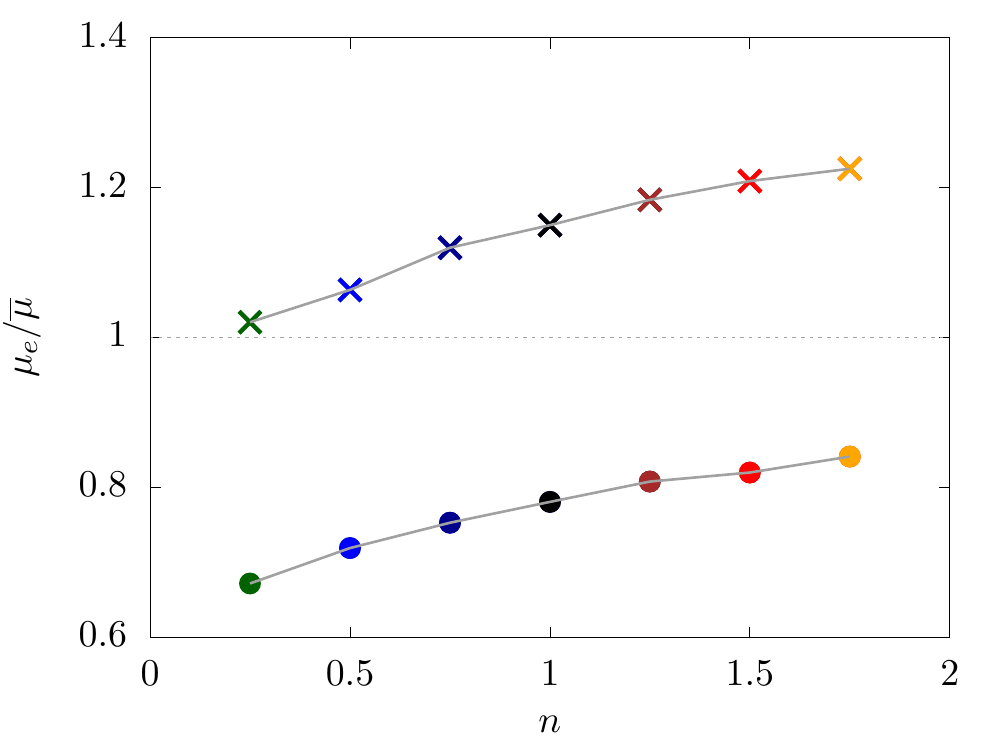}
  \includegraphics[width=0.45\textwidth]{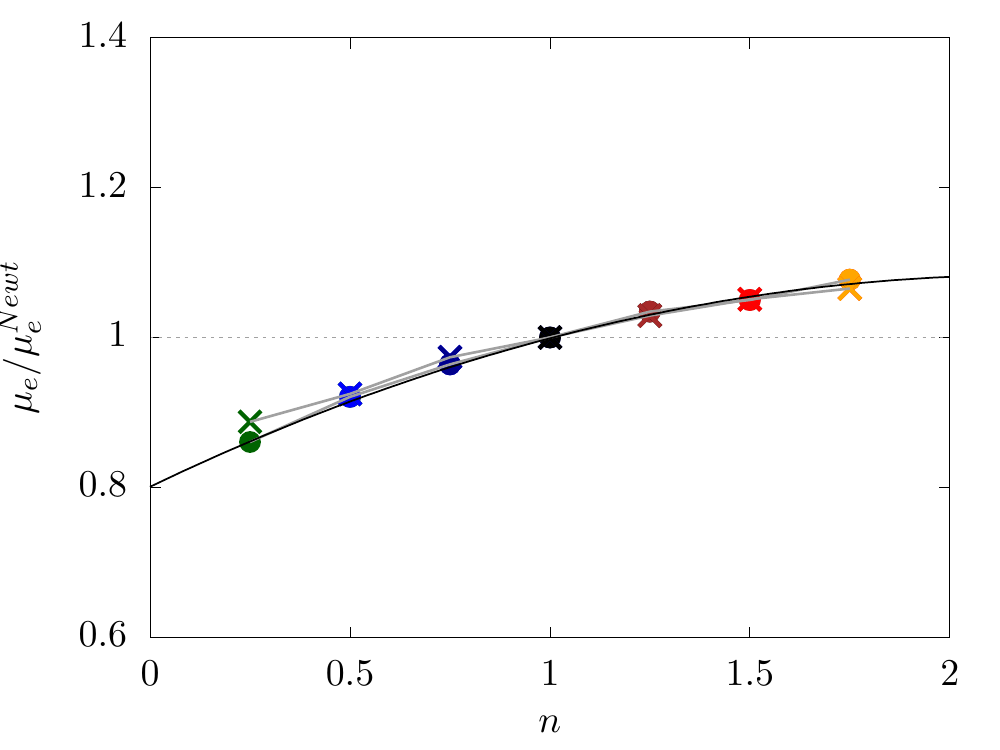}
  \includegraphics[width=0.45\textwidth]{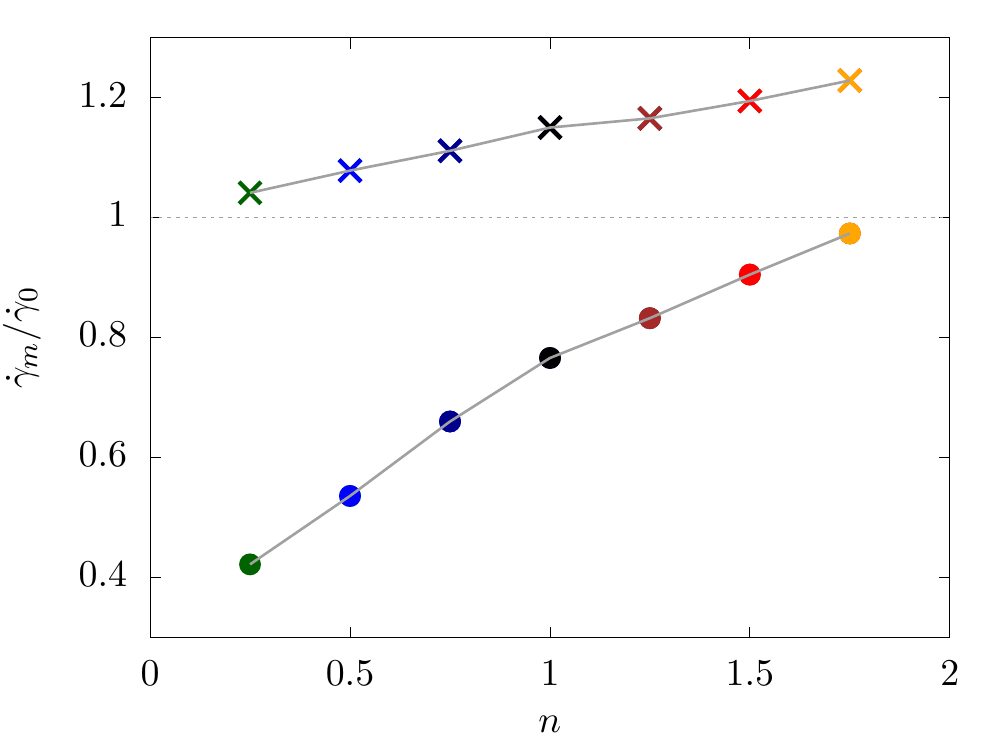}
  \includegraphics[width=0.45\textwidth]{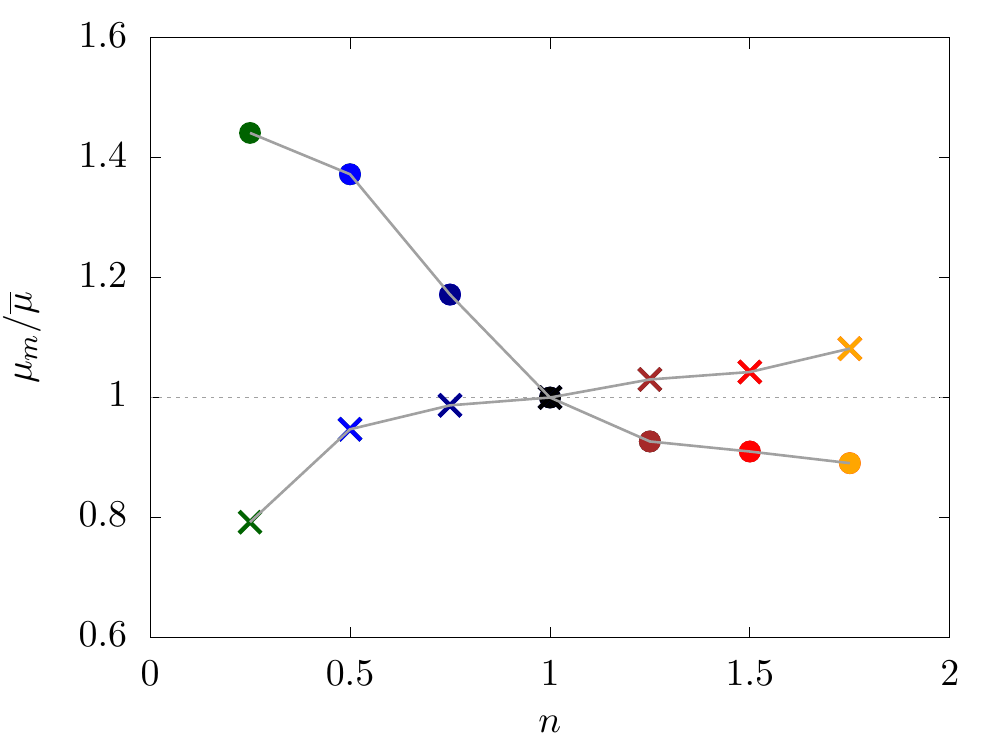}
  \caption{(top left) Normalised effective viscosity $\mu_e$, (top right) ratio of the effective viscosity $\mu_e$ with the Newtonian effective viscosity $\mu_e^{Newt}$ (bottom left) \mr{mean shear rate $\dot{\gamma}_m$ and (bottom right) viscosity $\mu_m$} as a function of the flow index $n$. All cases are at $\Phi=0.2$ and the cross $\pmb \times$ and circle $\bullet$ symbols are used to distinguish the cases with $\lambda$ equal to $1$ and $0.01$, respectively. The black line in the top right figure is a numerical fit to the data in the form of $\mu_e = \mu_e^{Newt} \left( 0.801+0.256 n-0.058 n^2 \right)
$.}
  \label{fig:n}
\end{figure*}
Finally, we analyse in more details how the results changes with the power index $n$. The top left panel of \figrefS{fig:n} reports the effective viscosity $\mu_e$ as a function of $n$ for the two viscosity ratio $\lambda$ investigated and for a fixed volume fraction $\Phi=0.2$. The figure shows that $\mu_e$ grows monotonically with $n$ for both the viscosity ratio tested. Furthermore, the figure also suggests that the rate of change of the effective viscosity is approximately independent of $\lambda$. This is proved in the top right panel of the figure, where we report the effective viscosity $\mu_e$ divided by the corresponding effective viscosity of the Newtonian case $\mu_e^{Newt}$ (\ie $n=1$). The two curves obtained for different viscosity ratio $\lambda$ well collapse into a single curve, whose expression can be obtained with a simple numerical fit as
\begin{equation} \label{eq:fit}
\mu_e = \mu_e^{Newt} \left( 0.801+0.256 n-0.058 n^2 \right).
\end{equation}
The fit is also reported in the figure with a black line showing a good level of agreement with the results of our simulations. This result provides the important conclusion that the effective viscosity of an emulsion in a power law fluid can be well predicted with \equref{eq:fit} knowing only the effective viscosity for the Newtonian configuration. When we study the \mr{mean values of the shear rate $\dot{\gamma}_m$ and of the viscosity $\mu_m$} the result is quite different. In particular, while the \mr{mean shear rate $\dot{\gamma}_m$} still grows monotonically with $n$, its rate of change is enhanced by lower values of the viscosity ratio $\lambda$. This is a consequence of the enhanced coalesce of the droplets arising from the combined effect of the low viscosity ratio and of the shear-thinning property of the fluid. The \mr{mean viscosity $\mu_m$} increases with $n$ for the unit viscosity ratio case $\lambda=1$ while it decreases for the case with $\lambda=0.01$. This results further shows that the largest variations of \mr{mean viscosity} are found in the case with low viscosity ratio $\lambda$ while much smaller variations are evident as $\lambda$ grows.

\begin{figure}
  \centering
  \includegraphics[width=0.45\textwidth]{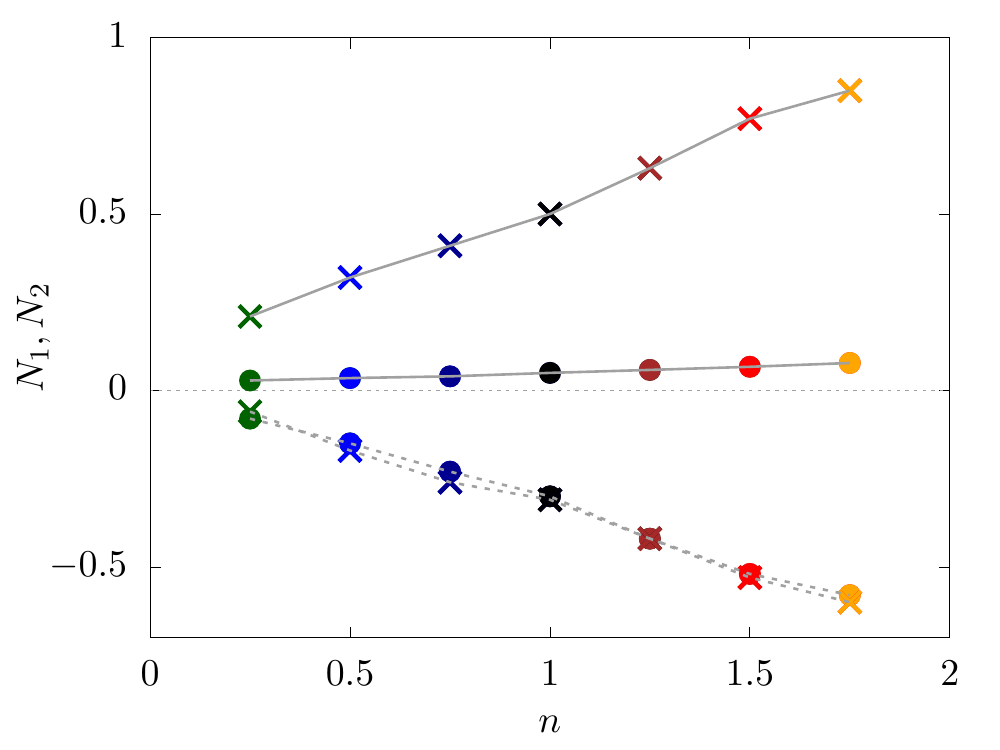}
  \caption{Normalised first and second normal stress difference $\mathcal{N}_1$ and $\mathcal{N}_2$ as a function of the flow index $n$. All cases are at $\Phi=0.2$ and the cross $\pmb \times$ and circle $\bullet$ symbols are used to distinguish the cases with $\lambda$ equal to $1$ and $0.01$, respectively.}
  \label{fig:n_n}
\end{figure}
We conclude our investigation by showing in \figrefS{fig:n_n} the first and second normal stress differences of the system, defined as
\begin{subequations}
\begin{align}
\mathcal{N}_1 &=\frac{\langle \sigma_{11} -\sigma_{22} \rangle}{ \overline{\mu} \dot{\gamma}_0}, \\
\mathcal{N}_3 &=\frac{\langle \sigma_{22} -\sigma_{33} \rangle}{ \overline{\mu} \dot{\gamma}_0}.
\end{align}
\end{subequations}
From the figure we observe that the magnitude of both the normal stress differences grows with the flow index $n$. The first normal stress difference $\mathcal{N}_1$ is greater then zero, \ie $\sigma_{11} > \sigma_{22}$, and thus droplets are elongated in the direction of the flow and compressed in the wall-normal direction. On the other hand, the second normal difference $\mathcal{N}_2$ is always negative. While $\mathcal{N}_1$ substantially reduces for $\lambda=0.01$, $\mathcal{N}_2$ is mostly unaffected by the viscosity ratio.

\section{Conclusions} \label{sec:conclusion}
We study the rheology of a two-fluid emulsion in semi-concentrated conditions with the solute being a Newtonian fluid and the solvent an inelastic power law fluid. The problem is studied numerically by the volume of fluid method. Different volume fractions and viscosity ratios are considered together with several carrier fluids exhibiting both shear-thinning and thickening behaviours.  We do not vary the capillary number which is fixed to a value such that a single droplet subject to the applied shear rate does not breakup, in order to focus our attention on the droplet coalescence mechanism.

First, we study how the effective viscosity of the system changes with the different carrier fluid properties. We show that the effective viscosity grows with the volume fraction of the dilute phase, reaches a maximum and then decreases with it beyond a certain concentration; this critical concentration corresponding to the maximum effective viscosity reduces when decreasing the viscosity ratio. The general behaviour remains mostly unaltered when considering power law fluids, except for an increase in effective viscosity for shear-thickening fluids and a decrease for the shear-thinning ones.

We show that the \mr{mean shear rate} exhibit a behaviour similar to the effective viscosity, while the \mr{mean viscosity} does not. Indeed, the mean viscosity is less than the nominal one for shear-thinning fluids and larger than the nominal one for shear-thickening fluids with unit viscosity ratios and the opposite when the viscosity ratio is reduced. This is consistent with the values of local shear rates measured in the domain but in contrast with the value of the effective viscosity. We explain this contradiction by studying the coalescence efficiency of the system which is modified by the nature of the carrier fluid. To support our claim, we show that the local shear rate assumes a very wide range of values; although its probability density function is peaked at the mean value, the distribution is strongly skewed exhibiting strong tails especially for large shear rates. This local high shear rate corresponds to the regions in between two interacting droplets, thus increases with the solute volume fraction. As a consequence, local high and low viscosity arises in these regions for shear-thickening and thinning fluids, respectively, which ultimately affect the coalescence of the droplets. Indeed, we relate the changes in the emulsion viscosity mainly to modifications of the coalescence in the system obtained by changing the carrier fluid property: coalescence is enhanced for shear-thinning fluids and reduced for shear-thickening ones due to modifications of the drainage time of the system, caused by modifications of the viscosity in the system. This process is mainly dominated by the power law fluid index $n$, with shear-thinning and thickening fluids exhibiting two different behaviours which are (qualitatively) independent of the nominal viscosity ratio of the two fluids; also, this can not be understood by considering only the mean shear rate and viscosity of the two fluids across the domain.

Finally, we provide an expression able to successfully estimate the effective viscosity of the emulsion, as a function of power law index $n$ and of the effective viscosity of the emulsion in a Newtonian fluid at the same volume fraction and viscosity ratio. This is a consequence of the fact that $\mu_e$ grows with $n$ with a rate of change which is approximately independent of $\lambda$.

To conclude, our results show that the coalescence efficiency which strongly affects the system rheology can be controlled by properly choosing the non-Newtonian property of the carrier fluids. Fully neglecting droplet merging can lead to erroneous or incomplete predictions in several flow conditions. In this work we have investigated a limiting cases where the nominal coalescence efficiency is close to unity; in a real scenario, the coalescence efficiency is likely to have intermediate values and future works may therefore be devoted to handling both coalescence and collisions in order to simulate a more wide range of emulsions. Anyway, the present choice helped to highlight the importance of coalescence.

\appendix*
\section{Verification of the numerical method and of the results}
\begin{figure}
  \centering
  \includegraphics[width=0.45\textwidth]{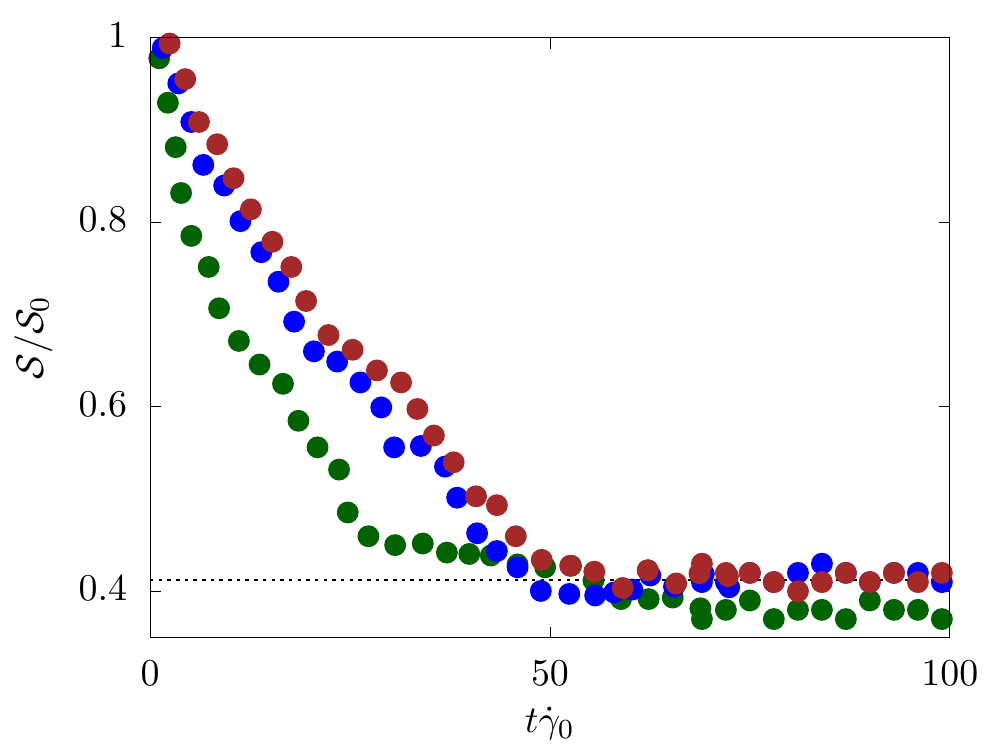}
  \caption{The total surface area $\mathcal{S}$ normalised with its initial value $\mathcal{S}_0$ as a function of the non-dimensional time. The trhee colors are used to indicate different grid sizes: green, blue and brown correspond to $8$, $16$ and $32$ grid points per initial droplet radius $\mathcal{R}$.}
  \label{fig:St}
\end{figure}
When two droplets approach each other, they may coalesce or not, with the coalescence happening without the addition of any non-hydrodynamic force due to the finite (although small) inertia. Topological changes of droplets such as coalescence and breakup are treated in the present methodology automatically without any additional model, since with the volume of fluid method these happen automatically when the distance between two interface falls within a grid cell. However, this does not mean that the coalescence process is correctly captured. On the other hand, fully resolving the coalesce is numerically challenging and not feasible, due to the very small scales involved, which are order of magnitude smaller than the characteristic size of the droplets. Notwithstanding the fact that coalescing events are here numerically driven, due to the finite size of our grid which will cause droplets to merge, we can verify that the present simulations still provides meaningful macroscopic results. To do that, we have proceeded as follows: first, we use the same setup employed in a previous work where the rheological data of a Newtonian droplet suspension have been compared with available experimental measurements \citep{de-vita_rosti_caserta_brandt_2019a}; next, we have verified that our grid resolutions is fine enough to provide results which are weakly dependent on the grid size resolution, by evaluating their changes when doubling the grid size, and found variations less than $5\%$. Note that, while this ensure that the bulk results are accurate, it does not mean that we are fully capturing the very small dynamics corresponding to the actual drainage process.

To prove this, \figrefS{fig:St} reports the time evolution of the total surface area for a case with $\Phi=10\%$ and viscosity ratio $\lambda=0.01$. This test was chosen because the surface area is the quantity that is mostly affected by coalescence, being its changes directly related to it, and is thus a good indicator to show if coalescence is properly resolved; also the case chosen has a low viscosity ratio which is the one where coalescence is promoted, thus being the most demanding in terms of resolution. We can observe that after an initial transient phase where the surface area decreases, \ie droplets are coalescing, the system reaches a statistically steady state where the surface area slightly oscillates around a mean value, indicated by the dashed line and extracted by time averaging over the second half of the time signal. The figure report the same quantity obtained using three different grid sizes, $8$, $16$ and $32$ grid points per initial droplet radius $\mathcal{R}$ (corresponding to $80$, $160$ and $320$ grid points per channel height $2h$). We can observe that, although the time-histories are different,  with the coarse grid leading to a faster coalescence than the finer ones, the steady state value obtained with the two finer grids are comparable, while a slightly lower value is obtained for the coarsest grid. The figure suggests that the grid used in the present simulations ($16$ grid points per initial droplet radius $\mathcal{R}$) is appropriate to properly describe the macroscopic phenomenon under investigation.

We have also similarly tested that the results are independent of the size of the homogeneous directions $x$ and $z$ and that the wall-to-wall distance guarantees a low level of confinement similarly to what found in previous works \citep{fornari_brandt_chaudhuri_lopez_mitra_picano_2016a, takeishi_rosti_imai_wada_brandt_2019a}.

\section*{Acknowledgments}
MER was supported by the FY2019 JSPS Postdoctoral Fellowship for Research in Japan (Standard) P19054 and by the JSPS KAKENHI Grant Number JP20K22402. The authors acknowledge computer time provided by the Supercomputing Division of the Information Technology Center, The University of Tokyo, and by the Scientific Computing section of Research Support Division at OIST.

\section*{Data availability}
The data that support the findings of this study are available from the corresponding author upon reasonable request.

%\bibliography{../../../Articles/bibliography.bib}

\end{document}